\documentclass[journal]{IEEEtran}

\usepackage{amsmath,amsfonts,amssymb,mathrsfs}
\usepackage{graphicx}
\usepackage{setspace}
\usepackage{tocloft}
\usepackage{url}
\usepackage{hyperref}
\usepackage{booktabs}
\usepackage{xcolor}
\usepackage[capitalize]{cleveref}
\usepackage{bm}
\usepackage{cite}

\newcommand{\tred}[1]{#1}

\begin{document}

\title{On the unreasonable effectiveness of CNNs}

\author{Andreas Hauptmann,~\IEEEmembership{Member,~IEEE,}
and Jonas Adler
\thanks{This work was partially supported by the Academy of Finland Project 312123 (Finnish Centre of Excellence in Inverse Modelling and Imaging, 2018--2025) and the CMIC-EPSRC platform grant (EP/M020533/1)}%
\thanks{A. Hauptmann is with the Research Unit of Mathematical Sciences; University of Oulu, Oulu, Finland and with the Department of Computer Science; University College London, London, United Kingdom.}%
\thanks{J. Adler did this work at the department of mathematics; KTH -- Royal Institute of Technology, Stockholm Sweden. He is currently with DeepMind, London, UK.}%
}%

\maketitle

\begin{abstract}
	Deep learning methods using convolutional neural networks (CNN) have been successfully applied to virtually all imaging problems, and particularly in image reconstruction tasks with ill-posed and complicated imaging models.
	In an attempt to put upper bounds on the capability of baseline CNNs for solving image-to-image problems we applied a widely used standard off-the-shelf network architecture (U-Net) to the ``inverse problem'' of XOR decryption from noisy data and show acceptable results.
\end{abstract}

\section{Introduction}

An ever-increasing amount of data and emerging methods in deep learning have led to considerable advances in numerous computer vision tasks, such as object detection and classification. The underlying methodology is based on convolutional neural networks (CNN), which can be understood as a multi-layered feature extraction from neighbourhood relations in the input image.

The subclass of methods that we consider here are image-to-image networks and in particular, we are motivated by their immense impact in inverse problems and biomedical imaging applications.
This success is partly driven by the availability of large amounts of data and the tendency to open-source this information for other scientists, but also by the ability to learn visually appealing and data specific representations. Consequently, we are now experiencing a transition to utilise tools from data science to harness the potential of large data, which is in stark contrast to classical deterministic algorithms that only operated with few data. For instance, whereas in tomographic imaging meticulously tuned algorithms were developed for reconstructions, we can now simply train a network to recover the important features in a tomographic image, if sufficient data is available.

This transition is marked in parts due to the success of 
\tred{widely established} convolutional neural network \tred{architectures,} such as the U-net which was initially used for semantic segmentation \cite{ronneberger2015u}, but has since been utilised for various imaging tasks as an essential processing step to improve image quality. For instance, to remove noise in low-dose CT images \cite{kang2017deep,jin2017deep} or artefacts from undersampled magnetic resonance imaging \cite{lee2017deep,hauptmann2019real}, 
and as essential part in the pipeline for automated diagnosis \cite{de2018clinically}.

\tred{However, it is not unusual to propose a new method using different network designs \cite{zhu2018image,yang2017dagan}, or intertwine learned components with explicit hand-designed operations \cite{hammernik2018learning,qin2018convolutional, mardani2018deep}. These methods generally provide impressive results, but it is often claimed without computational proof that the problem under consideration is too special for the application of basic network architectures. We believe, this is in part due to the common understanding that the problem has to be sufficiently well-behaved for standard CNNs to be applicable \cite{arridge2019solving}.} For example, the use of convolutions would imply that the underlying problem should be translation equivariant\footnote{A function is translation equivariant if translating the input and then applying the function is equivalent to applying the function and then translating the output. All convolutions satisfy this property.}, we refer to \cite{adler2017solving} for a discussion in the context of inverse problems. Furthermore, the continuity and almost everywhere differentiability of the neural networks seemingly implies that the functions they approximate should at least be continuous \cite{hornik1991approximation}.

These somewhat contradictory trains of thought lead us to our question: \emph{are there really any image-to-image tasks that can not be solved reasonably well with a standard CNN and sufficient data?} Our main result is that even for seemingly ridiculous image-to-image problems standard CNNs basically always perform acceptably.
In particular we'll show that an off-the-shelf U-Net can be trained to invert XOR encryption, a function which is \emph{everywhere discontinuous} and \emph{not translation equivariant}. The results hold even in the noisy case where standard decryption fails.

\begin{figure*}[h]
\centering
\includegraphics[width=\textwidth]{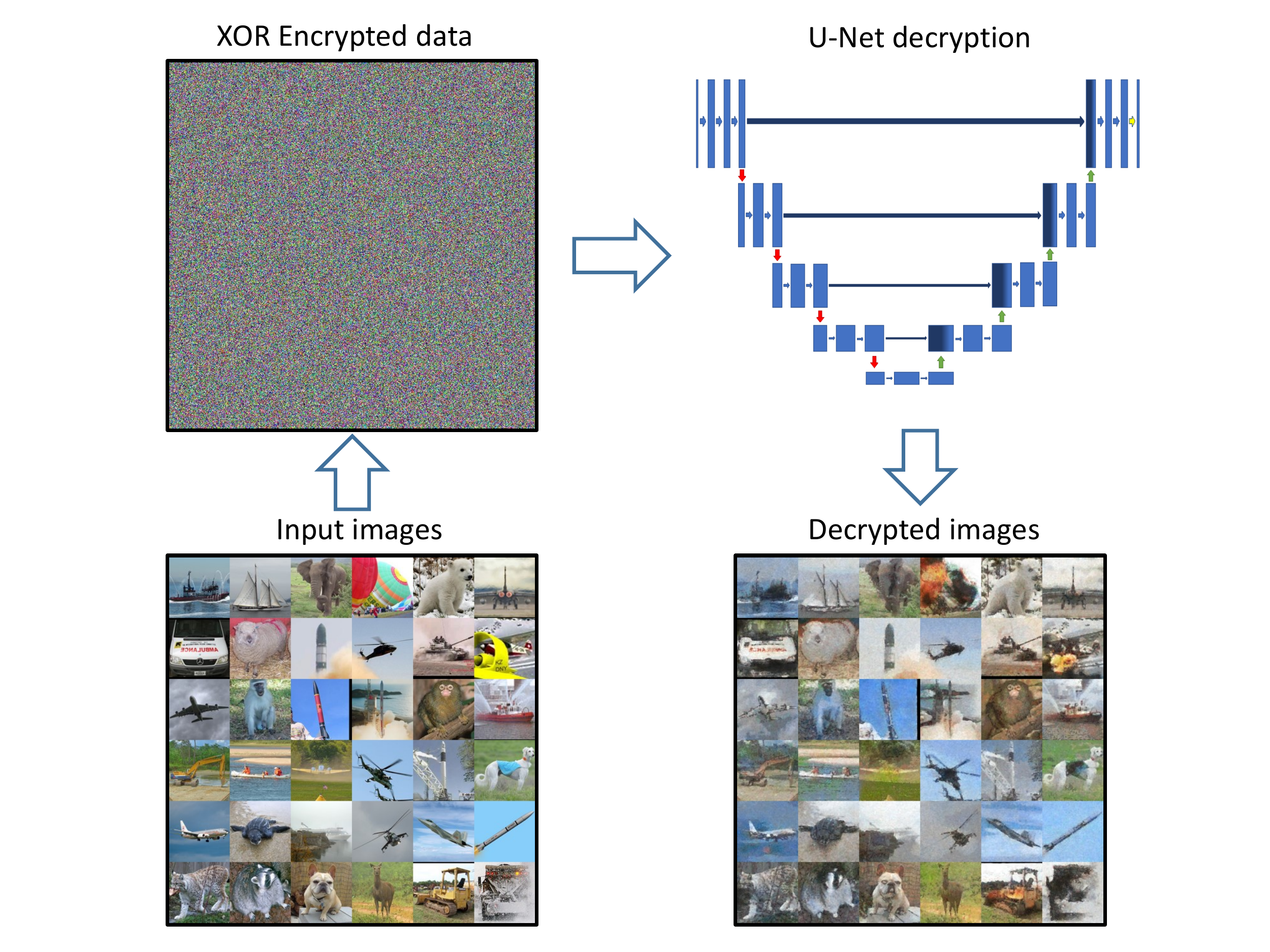}
\caption{\label{fig:diagram}Experimental setup for XOR encryption of natural images and decryption using a standard off-the-shelf deep convolutional neural network. The network is trained using known cipherimages (top-left images, 3 out of 12 channels shown here) $\Leftrightarrow$ plaintext (bottom left) pairs. Reconstructed test images after successful training (bottom right) exhibit a loss of resolution and detail, but retain identifying characteristics.} 

\end{figure*}




\section{Experimental setup}
\label{sec:Experiments}

We shall consider the inverse problem of inverting XOR encryption. Our forward operator is hence the XOR operation with a fixed byte string and the implementation we use is based on running the advanced encryption standard (AES) \cite{nechvatal2001report} in Counter mode (CTR) \cite{diffie1979privacy}, \tred{a block cipher with randomly generated initialisation (iv) and key, here 128 bit long and fixed for all examples.}
The encryption process first generates a string of bytes using the initialisation and key and then takes the input, represented as a sequence of bytes, applies a bitwise XOR of the input and byte string, and then returns a byte array of 
the same size, called the ciphertext. The recipient of the ciphertext can then use the key in order to recover the input. For obvious reasons, the process has been designed to be highly discontinuous in the input in order to thwart attackers from reading the text. To make the problem even harder, we also study the case with noisy observations.

As with any standard machine learning method for inverse problems, the recovery of the encrypted images can be formulated as a basic supervised learning problem. That is, given a set of ground-truth images $\{f_i\}$ with corresponding measured data $\{g_i\}$, we then formulate a network $\Lambda_\theta$ with parameters $\theta$ to recover the ground-truth from the encrypted image, such that $\Lambda_\theta(g_i) \approx f_i$. This corresponds to the standard procedure in image processing, where $g_i$ represents a corrupted image, e.g. obtained from undersampled data and/or under high noise or even, as in our case, from bit-wise encoding. The training is then given as optimisation problem to find an optimal set of parameters $\theta^*$ by minimising a loss functions such as
\begin{equation}\label{eqn:loss}
\theta^* = \arg\min_\theta \frac{1}{N}\sum_{i=1}^N\|\Lambda_\theta(g_i) - f_i \|^2_2.
\end{equation}

For training, we use the standardised STL-10 dataset \cite{coates2011analysis}, a set of 100,000 natural 96x96 pixel colour images (bottom-left of \cref{fig:diagram}).
The images we seek to encode are RGB images\tred{, which we normalised to the range $[0,1]$ and} then stored as single precision floating point numbers, three per pixel (one per colour channel). 
We encrypt the raw byte-representation of the image using AES in CTR mode with the fixed 128 bit key and fixed iv. \tred{As we aim to establish an image-to-image problem, we need to reinterpret the ciphertext as an image and make it admissible as input to the U-Net. To achieve this, we first converted the encryption output to \texttt{uint8}, then normalised similarly to the ground-truth images and reformatted to a $96 \times 96$ image with resulting $12$ channels in single precision floats,} which we call the call cipherimages, see top-left of \cref{fig:diagram} for examples. In the noisy setting we add 1\% pixel- and channel-wise Gaussian noise.


We split the data into 90,000 training images and 10,000 images for testing. The corresponding encrypted images were then computed for both sets with the same settings and noise added. A standard U-net architecture as originally proposed \cite{ronneberger2015u}, with a minor modification for consistency in image size and linearly rescaling the inputs to [0, 1), was used. The training was done by minimising \cref{eqn:loss} and performed with standard choices on the hyperparameters, using the Adam optimiser, batch size of 32, learning rate $2\cdot10^{-3}$ and 35 epochs.

The XOR forward operator has a closed form inverse and this inverse serves as our decryption baseline. In the noisy case we first round to the closest byte. We also consider another trivial baseline to see if we have actually learned decryption and not just some statistics about the dataset. Here, we note that the minimise of \cref{eqn:loss} if $g_i$ is uninformative of $f_i$ is the mean of the training set, $\bar{f}=\frac{1}{N}\sum_{i=1}^N f_i$ and use this as the reconstruction for any input.

\section{Discussion of results}\label{sec:results}

\begin{figure}[t]
\centering
\includegraphics[width=0.45\textwidth]{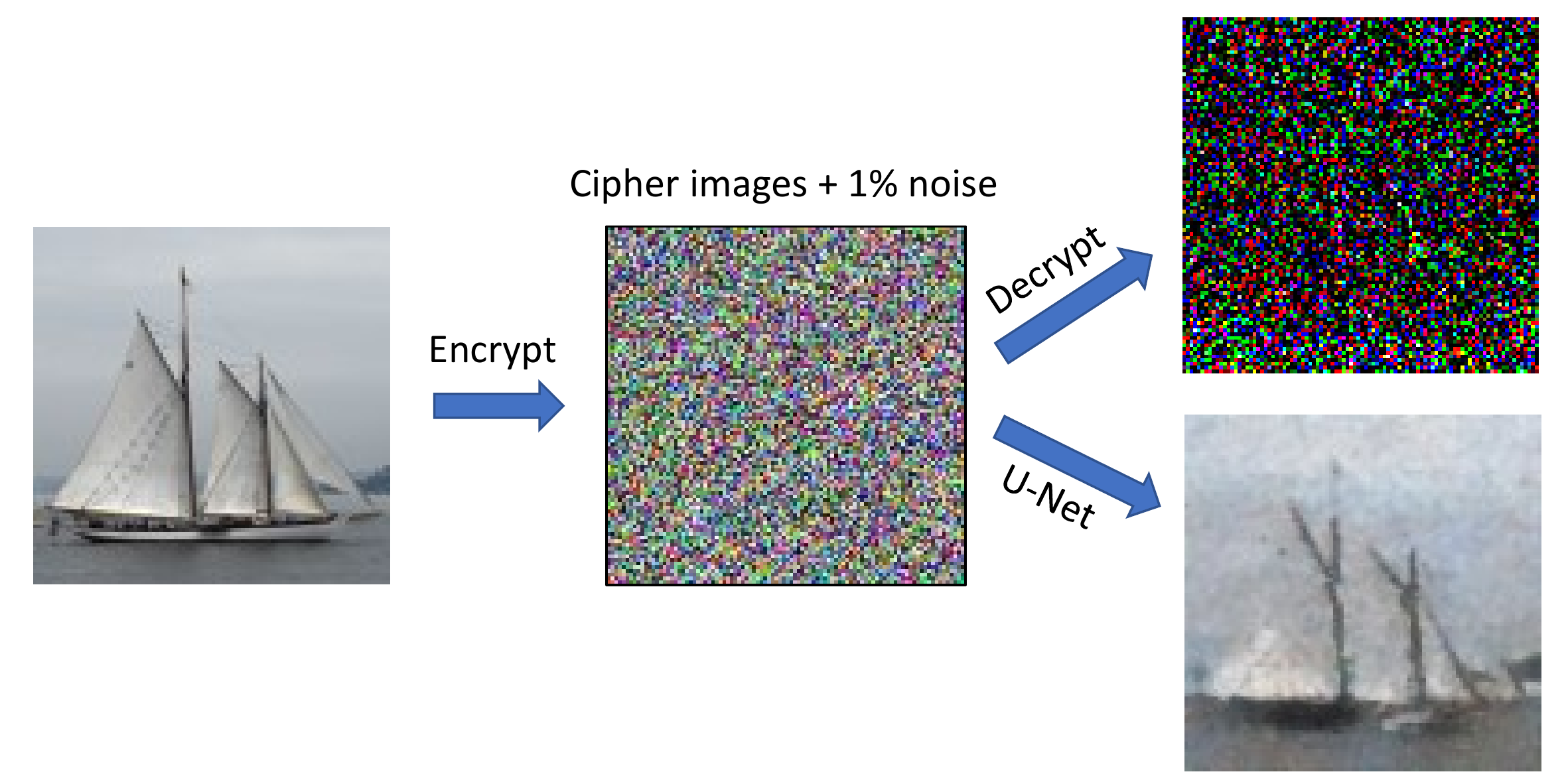}
\caption{\label{fig:noiseRecons} Reconstruction from corrupted data. We added 1\% noise to the cipherimages (Middle, shown 3 out of 12 channels). Resulting reconstructions on the right are with the XOR decryption with known key (top) and learned reconstruction (bottom). } 

\end{figure}

\begin{table}[h!] 
\small
  \caption{Quantitative values for the recovered test images. Mean values for 10,000 test samples.} 
  \begin{center}
    \begin{tabular}{l|r|r|r|r}
    & \multicolumn{2}{c|}{No noise} & \multicolumn{2}{c}{1\% Noise}\\
    &{\sc PSNR} &{\sc SSIM} &{\sc PSNR} &{\sc SSIM}   \\
    \toprule
  {\sc U-Net } &   18.0  &     0.62  &   \textbf{17.6}  &     \textbf{0.57}\\ 
  {\sc Decryption } &   $\bm{+\infty}$  &    \textbf{1.00}  &   $-\infty$  &  0.00\\ 
  {\sc Mean of train set } &   11.8  &    0.16  &  11.8 & 0.16\\ 
    \bottomrule
    \end{tabular}%
  \label{table:Quantitative}%
  \end{center}
\end{table}%

To our surprise, the network can successfully establish a relation between the cipherimages and the original input images. A set of such reconstructed images by the network from the test set is shown in \cref{fig:diagram}. As one can see, the reconstructed images suffer from a loss of resolution and some colours are not correctly recovered. Nevertheless, we can say that the reconstruction quality is well beyond our expectation.

This qualitative observation is also supported in terms of quantitative values, as the mean PSNR of all 10,000 reconstructed test images after training was 18dB and a SSIM of 0.62, as shown in \cref{table:Quantitative}.  In this noise free setting the baseline decryption method performs perfectly and obtains an infinite PSNR, as expected.
In the noisy setting with the learned decryption we observe a minor deterioration in mean PSNR of 0.4dB and 0.05 in SSIM,  whereas decryption from the corrupted data is impossible, giving a PSNR of negative infinity. We illustrate this in \cref{fig:noiseRecons}. 
Additionally, the simple mean baseline gives notably worse results, indicating that a nontrivial relation has indeed been successfully established.


As for limitations of this study, 
we note that while our encryption operation is different for each pixel (due to the XOR key being different) and hence not translation equivariant, there is no connection between pixels. We also tried running AES in Cipher Block Chaining (CBC) mode, which introduces a strong dependency, but were unable to learn anything useful in this setting. This was also the case for lower-dimensional training data such as the MNIST dataset.

Finally, we reiterate that a safe encryption uses varying keys as well as a random iv. If these are fixed, the encryption is deterministic and consequently not safe. This work does not show that supervised learning can crack encryption, what it does show is that supervised learning can solve problems with terrible numerical properties.

\section{Conclusions}\label{sec:conclusions}

Our results indicate, that against ones intuition, an off-the-shelf CNN can successfully learn a relation even in this extremely challenging setting. We believe there are at least two important takeaways from this observation that should be kept in mind when solving image-to-image problems:
\begin{itemize}
    \item With enough data, basic CNNs such as a U-Net, always work on image-to-image problems.
    \item Training a simple CNN should always be used as a strong baseline for \emph{any} newly proposed method, regardless of how unreasonable it seems.
\end{itemize}

\section*{Acknowledgement}
The authors would like to thank Ozan \"Oktem for suggesting that decryption ought to be an impossible inverse problem for supervised deep learning. We also thank Sebastian Lunz and Olaf Ronneberger for helpful discussions and comments.

\bibliographystyle{unsrt}
\bibliography{paperRefs}

\end{document}